\documentclass[dvips,12pt]{article}

\usepackage{color,graphics,palatino,fancyhdr}

\newcommand {\half} {{1 \over 2}}
\newcommand {\hhalf} {{\textstyle{1 \over 2}}}
\newcommand {\hhhalf} {{\scriptstyle{1 \over 2}}}

\newcommand {\ket}[1] {\left| #1 \right>}
\newcommand {\braket}[2] {\left< #1 | #2 \right>}
\newcommand {\Tr} {\mathrm{Tr}\,}
\newcommand {\diag} {\mathrm{diag}\,}
\newcommand {\im} {\mathrm{im}\,}
\newcommand {\sutwo} {{su}(2)}
\newcommand {\SUTWO} {\mathrm {SU}(2)}

\begin{document}

    \title{
            {\footnotesize{\hfill BROWN-HET-1282}}
            \\Ghosts as Negative Spinors}
    \author{Andr\'e van Tonder
            \thanks{Work supported in part by DOE grant number DE FG02-91ER40688-Task A}
            \\ \\
            Department of Physics, Brown University \\
            Box 1843, Providence, RI 02906 \\
            andre@het.brown.edu}
    \date{July 10, 2002}

    \maketitle

    \begin{abstract}
        We study the the properties of a BRST ghost degree of freedom
        complementary to a two-state spinor.

        We show that the ghost may be regarded as a unit carrier of
        negative entropy.

        We construct an irreducible
        representation of the $\sutwo$ Lie algebra with negative
        spin, equal to
        $-\half$,
        on the ghost state space
        and discuss the representation of finite $\SUTWO$ group
        elements.

        The Casimir operator $\mathbf{J}^2$ of the combined spinor-ghost
        system is nilpotent and coincides with the BRST operator $Q$.
        Using this, we discuss the sense in which the positive and negative spin
        representations cancel in the product to
        give an effectively trivial representation.  We compute an
        effective dimension, equal to $\half$,
        and character for the ghost representation
        and argue that these are consistent with this cancellation.
    \end{abstract}

    \section{Introduction}

    The technique of using ghosts to compensate
    unphysical degrees of freedom in the covariant
    quantization of systems with constraints was
    originally
    introduced by Faddev and Popov \cite{FP}.  Tyutin, Becchi, Rouet
    and Stora \cite{BRST} discovered that the ghost-enhanced system
    is invariant under a nilpotent (BRST) symmetry.  This symmetry
    can be used to
    identify physical states and operators using the language of cohomology
    \cite{HENNEAUX}.

    Ghosts give mathematical realization to
    the concept of negative degrees of freedom, which carry negative entropy
    or information.  Although originally introduced
    as a mathematical device, these negative degrees of freedom are
    interesting to study in their
    own right.

    In this article we will concentrate on
    the most elementary negative degree of freedom -- a BRST ghost
    complementary to a single two-state spinor.

    Since we will show that the ghost entropy
    ranges between $0$ and $-1$ and cancels the entropy of the
    spinor, the ghost may be regarded as the negative version of
    a quantum bit, carrying $-1$ bit of information.

    We then construct an irreducible
    representation of the $\sutwo$ Lie algebra with negative spin,
    equal to $-\half$,
    on the ghost state space.
    We also discuss the representation
    of finite $\SUTWO$ group elements.  While we are able to
    write down the matrix representation of a restricted class of
    group elements,  we find that
    the Lie algebra representation does not exponentiate to a
    representation of the full group.

    The
    Casimir operator $\mathbf{J}^2$ of the combined spinor-ghost
    system is nilpotent and coincides with the BRST operator $Q$.
    Using this, we discuss the sense in which the positive and negative spin
    representations cancel in the product to
    give an effectively trivial representation.  We compute an
    effective dimension, equal to $\half$, and character for the ghost representation
    and argue that these are consistent with this cancellation.

    \section{The ghost}

    The definition of
    the ghost complementary to a two-state spinor may be motivated as
    follows.

    Consider a Hamiltonian on the spinor state space of the form
    \[
        H_s = \omega \left(
                    \begin{array}{rr}
                        -\hhalf &  \\  & \hhalf
                    \end{array}
            \right)
    \]
    in a basis which we denote by $\{\ket{\scriptstyle{-\half}}_s, \ket{\hhhalf}_s\}$.
    The spinor partition function at finite inverse temperature $\beta$ is
    given by
    \[
        Z_s (\beta) = \mathrm {Tr}\, e^{-\beta H_s} = q^{-\half}(1 + q),\qquad
        q\equiv e^{-\omega\beta}.
    \]
    We would like to introduce a ghost degree of freedom
    that will cancel the contribution of the spinor to
    the partition function.
    Assuming that the spinor and the ghost do not interact,
    the combined partition function $Z(\beta)$ will factorize
    \[
        Z(\beta) = Z_s(\beta) \, Z_{g} (\beta).
    \]
    and the ghost partition function should be
    \[
        Z_{g}(\beta) = {q^{\half} \over 1 + q}.
    \]
    Expanding this as a geometric series
    \begin{equation}
        Z_g = q^{\half} \left(1 - q + q^2 - q^3 + \cdots\right),
        \label{ALTERNATE}
    \end{equation}
    we can write the ghost partition function in the form
    \begin{equation}
        Z_g = \Tr \eta\, e^{-\beta H_g},
         \label{TWISTEDPARTITION}
    \end{equation}
    where the matrix in the exponent
    \begin{equation}
        H_{g} = \omega\,\mathrm{diag}\,\left(\hhalf, \textstyle{3\over 2},
          \textstyle{5\over 2}, \ldots\right),  \label{HGHOST}
    \end{equation}
    defines a Hamiltonian on an infinite-dimensional vector space
    with ordered basis denoted by
    \[
    \{\ket{\hhalf}_g, \ket{\textstyle{3\over 2}}_g, \ket{\textstyle{5\over 2}}_g, \ldots\}
    \]
    and
    \[
        \eta = \mathrm{diag}\,(1, -1, 1, -1, \ldots).
    \]
    The presence of $\eta$ in the partition function may be motivated
    as follows.  Writing the partition function of the combined
    spinor-ghost system as
    \[
        1 = (1 + q)\, (1 - q + q^2 - \cdots) = 1 + q - q + q^2 - q^2 + \cdots,
    \]
    it is clear that the minus signs provided by $\eta$
    cause the contributions of excited states to
    cancel pairwise, leaving only the contribution of the ground
    state.  In the next section we will see that
    the ground state is the only physical state in
    the combined system.  Since the partition function should only
    count physical states, we conclude that $\eta$ is necessary.

    \section{BRST cohomology}

    In BRST theories \cite{HENNEAUX}, the analysis of physical states and
    operators is carried out in terms of an operator $Q$ that is hermitian
    and nilpotent.  In other words,
    \[
        Q^\dagger = Q, \qquad Q^2 = 0.
    \]
    Physical states satisfy
    \[
        Q\ket{\phi} = 0,
    \]
    and are regarded as equivalent if they differ by a $Q$-exact
    state. In other words,
    \[
        \ket{\phi} \sim \ket{\phi} + Q \ket{\chi},
    \]
    where $\ket{\chi}$ is an arbitrary state.
    More formally, physical states are elements of the
    cohomology of $Q$, defined as the quotient vector space
    \[
        {\ker Q\, / \,\im Q}.
    \]
    The inner product on this quotient space may be defined in
    terms of the original inner product by
    noting that all elements of $\im Q$ are orthogonal to all
    elements of $\ker Q$, so that the induced inner product
    defined on equivalence classes $[\ket{\phi}] = \ket{\phi} + \im Q$
    in the cohomology by
    \[
        \braket {\phi + \im Q}{\phi' + \im Q} \equiv
        \braket{\phi}{\phi'}, \qquad \phi, \phi' \in \ker Q
    \]
    is well defined.

    A hermitian operator $A$ is regarded as physical if $[A, Q] =
    0$.  This ensures that $A$ leaves $\im Q$ invariant, so that
    the reduced operator $[A]$ defined on the cohomology classes by
    \begin{equation}
        [A]\,\left(\ket{\phi} + \im Q\right) = A\ket{\phi} + \im Q  \label{AREDUCED}
    \end{equation}
    is well-defined.  In particular, the hamiltonian is required
    to be physical
    \[
        [Q, H] = 0.
    \]
    In order to define a BRST cohomology in the system
    at hand, we need to postulate a BRST operator $Q$.
    $Q$ should commute with the combined hamiltonian which,
    in the ordered basis
    \[
    \{\ket{\textstyle{-\half}}_s\otimes\ket{\hhalf}_g,
    \ket{\hhalf}_s\otimes\ket{\hhalf}_g,
    \ket{\textstyle{-\half}}_s\otimes\ket{\textstyle{3\over 2}}_g,
    \ket{\hhalf}_s\otimes\ket{\textstyle{3\over 2}}_g, \ldots \},
    \]
    has the form
    \[
        H = H_s + H_g = \left(
                \begin{array}{cccc}
                        0 &  &  &  \\
                          & \fbox{\ensuremath{\begin{array}{rr}
                                    1 &   \\
                                      & 1
                                  \end{array}}
                            }
                             &  & \\
                          &  &
                            \fbox{\ensuremath{\begin{array}{rr}
                                    2 &   \\
                                      & 2
                                  \end{array}}
                            }
                                &  \\
                          &  &  &  \ddots
                \end{array}
            \right).
    \]
    Thus, $Q$
    takes the two-dimensional subspace formed by each pair of excited states with
    the same energy eigenvalue onto itself.  Since the combined partition function
    $Z = 1$, the cohomology of $Q$ cannot
    contain excited states.  Therefore, $Q$ cannot be zero
    in any of the excited subspaces.  Since we need to have $Q^2 = 0$,
    we see that in each of these subspaces
    there has to be pair of linearly independent
    vectors $\ket{v_1}$ and $\ket{v_2}$ such
    that $Q\ket{v_1} = \ket{v_2}$ and $Q\ket{v_2} =
    0$.  This is precisely what one needs to eliminate these excited
    states from the cohomology.

    The operator $Q$ can only be hermitian if the inner product is
    not positive definite.  Indeed, assuming hermiticity,
    the inner product of
    $\ket{v_2}$ with itself is
    \[
        \braket {v_2}{v_2} = \braket {Q\, v_1}{Q\, v_1} = \braket {v_1} {Q^2 \, v_1} = 0.
    \]
    In other words, the vector $v_2$ is null (has zero norm).
    A suitable indefinite inner product is obtained by defining the
    inner product on the ghost state space
    as
    \begin{equation}
    \eta_{mn} \equiv \braket{m}{n}_g = (-)^{m-\half} \delta_{mn},
        \qquad m = \hhalf, {\textstyle {3\over 2}},
      {\textstyle {5\over 2}},\ldots,\label{INNERPROD}
    \end{equation}
    or
    \begin{equation}
        \eta = \diag (1, -1, 1, -1, \dots).
    \end{equation}
    In the product space, the inner product is then
    \[
       G \equiv \mathbf{1}_s \otimes \eta  = \left(
                \begin{array}{rrrrrrr}
                        1 &   &    &     &   &  &   \\
                          & 1 &    &     &   &  &   \\
                          &   & -1 &     &   &  &   \\
                          &   &    &  -1 &   &  &   \\
                          &   &    &     & 1 &  &   \\
                          &   &    &     &   & 1 &   \\
                          &   &    &     &   &   & \ddots
                \end{array}
            \right).
    \]
    Given this inner product, we will show in later sections that
    one can construct a hermitian representation
    of the $\sutwo$ Lie algebra on the state space.  In this
    representation, the Casimir operator $\mathbf{J}^2$ is nonzero but
    satisfies $(\mathbf{J}^2)^2 = 0$.  It is therefore a very natural
    candidate for $Q$, and we will indeed choose $Q = \mathbf{J}^2$.  This
    will have the added benefit that the $\sutwo$ generators will
    be physical operators in the sense discussed above.

    In the above basis, the operator $Q = \mathbf{J}^2$ has the form
    (see section 9)
    \begin{equation}
        Q = \left(
                \begin{array}{cccc}
                        0 &  &  &  \\
                          & \fbox{\ensuremath{\begin{array}{rr}
                                    -1 & i \\
                                    \phantom{-}i  & 1
                                  \end{array}}
                            }
                             &  & \\
                          &  &
                            \fbox{\ensuremath{\begin{array}{rr}
                                    -2 & 2i  \\
                                    \phantom{-}2i & 2
                                  \end{array}}
                            }
                                &  \\
                          &  &  &  \ddots
                \end{array}
            \right). \label{Q1}
    \end{equation}
    It is easily checked that $Q^2 = 0$ and that $Q$ is hermitian
    with respect to the inner product $G$, as follows from
    \[
        Q = Q^\dagger = G Q^+ G,
    \]
    where the dagger denotes hermitian conjugation with respect to
    the indefinite inner product $\braket{\cdot}{\cdot}$ on the state space and
    the $+$ sign denotes the usual matrix
    adjoint (\ref{PSEUDOHERMITIAN}).

    With respect to the basis consisting of the ground state and
    the null excited states
    \begin{eqnarray}
        \ket{n} &\equiv& {1\over \sqrt
        2}\left(i\,\ket{-\hhalf}_s\otimes\ket{n+\hhalf}_g
                    +
                    \ket{\hhalf}_s\otimes\ket{n-\hhalf}_g\right),
                    \nonumber\\
        \ket{\tilde n} &\equiv& {1\over \sqrt
        2}\left(-i\,\ket{-\hhalf}_s\otimes\ket{n+\hhalf}_g
                    +
                    \ket{\hhalf}_s\otimes\ket{n-\hhalf}_g\right),
                    \label{NTILDEN}
    \end{eqnarray}
    $G$ has the form
    \[
        \left(
                \begin{array}{cccc}
                        1 &  &  &  \\
                          & \fbox{\ensuremath{\begin{array}{rr}
                                      & 1  \\
                                    1 &
                                  \end{array}}
                            }
                             &  & \\
                          &  &
                            \fbox{\ensuremath{\begin{array}{rr}
                                      & -1  \\
                                    -1 &
                                  \end{array}}
                            }
                                &  \\
                          &  &  &  \ddots
                \end{array}
            \right),
    \]
    and $Q$ is
    \begin{equation}
           \left(
                \begin{array}{cccc}
                        0 &  &  &  \\
                          & \fbox{\ensuremath{\begin{array}{rr}
                                    0 & 2 \\
                                    0 & 0
                                  \end{array}}
                            }
                             &  & \\
                          &  &
                            \fbox{\ensuremath{\begin{array}{rr}
                                    0 & 4  \\
                                    0 & 0
                                  \end{array}}
                            }
                                &  \\
                          &  &  &  \ddots
                \end{array}
            \right). \label{Q2}
    \end{equation}
    By construction, the cohomology of $Q$ consists of the single
    zero-energy state
    \[
        \ket{\textstyle{-\half}}_s\otimes\ket{\hhalf}_g + \im Q.
    \]

    \section{Indefinite inner product spaces}

    We review a few facts regarding operators on indefinite inner
    product spaces such as (\ref{INNERPROD}), also
    called Kre\u{\i}n spaces in the
    literature \cite{INDEFINITE}.

    Unitary and hermitian operators on an indefinite product space are often
    called pseudo-unitary and pseudo-hermitian to emphasize that the
    inner product is indefinite.

    It is important to be aware that not all results
    that are valid for positive definite spaces are valid when the
    inner product is not positive definite.  For example, not all
    pseudo-hermitian operators are diagonalizable.  A good
    counterexample is precisely the operator $Q = \mathbf{J}^2$ above.

    In a basis where the metric tensor is represented by the matrix $\eta$,
    the matrix representation of a unitary transformation $U$
    satisfies
    \begin{equation}
        U^{-1} = U^\dagger = \eta \,U^+\, \eta,   \label{PSEUDOUNIT}
    \end{equation}
    where the dagger denotes hermitian conjugation with respect to
    the indefinite inner product and the plus sign denotes
    the ordinary matrix adjoint.

    A hermitian transformation satisfies
    \begin{equation}
        \mathcal{A} = A^\dagger = \eta \,\mathcal{A}^+\,\eta.
            \label{PSEUDOHERMITIAN}
    \end{equation}
    Taking $U = e^{i\epsilon A}$, it follows that the
    infinitesimal version of the pseudo-unitarity condition (\ref{PSEUDOUNIT})
    is just the pseudo-hermiticity condition
    (\ref{PSEUDOHERMITIAN}).

    \section{The oscillator representation}

    Now that we have determined the spectrum and the inner product
    on the ghost state space, we can write the partition function
    (\ref{TWISTEDPARTITION})
    in terms of a bosonic oscillator satisfying
    \begin{equation}
        [a, a^\dagger] = -1  \label{COMM}
    \end{equation}
    as
    \[
        Z_g = \Tr \eta\, e^{-\beta H_g},
    \]
    where
    \[
        H_g = -\omega\, a^\dagger a, \qquad \eta = (-)^N, \qquad N
        = - a^\dagger a.
    \]
    The minus sign on the right hand side of (\ref{COMM}) ensures
    that the inner product will have the indefinite form (\ref{INNERPROD}).
    We take the following normalization convention for the basis used in
    (\ref{TWISTEDPARTITION})
    \begin{equation}
      \ket{m}_g \equiv (-)^{m-\half}{1\over \sqrt {\left({m-\hhhalf}\right)!}}\,(a^\dagger)^{m-\hhhalf}
      \ket{\hhalf}_g, \qquad m = \hhalf, {\textstyle {3\over 2}},
      {\textstyle {5\over 2}}, \ldots \label {BASIS}
    \end{equation}

    \section{Entropy}

    From the expression
    (\ref{TWISTEDPARTITION}), we see that
    the finite temperature Boltzmann density
    matrix has the form
    \begin{eqnarray}
        \rho_g (\beta) &\equiv &{1\over Z_g}\,e^{-\beta H_g}\,\eta, \nonumber \\
             &=&
            (1 + q) \, \mathrm{diag}\,(1, -q, q^2, -q^3. \ldots),
            \qquad q = e^{-\beta}. \label{GHOSTDENSITY}
    \end{eqnarray}
    The density matrix has unit trace, as it should.  However, it has
    both positive and negative diagonal entries, so that the usual
    probabilistic interpretation is not valid if we regard the ghost in
    isolation.  However, it is possible to formally define an
    entropy for the ghost.

    Due to the negative eigenvalues in the ghost density matrix,
    we have to modify the conventional
    definition
    $S(\rho) \equiv - \Tr \rho\, \log \rho$
    of the von Neumann entropy \cite{NEUMANN}.  A suitable expression that
    is invariant under pseudo-unitary transformations
    $\rho \to U \rho U^{-1}$
    is
    \begin{equation}
        S (\rho) \equiv - \half\, \Tr \rho \log
        \left(\rho\right)^2.  \label{ENTROPYGHOST}
    \end{equation}
    This expression reduces to the von Neumann definition when the
    eigenvalues are positive.

    It is straightforward to verify that, with this definition,
    the entropy is additive for spinor-ghost states of the factorizable form
    $\rho =\rho_s\otimes \rho_g$.  In other words
    \[
        S(\rho_s\otimes\rho_g) = S_s (\rho_s) +
        S_g(\rho_g).
    \]
    Calculating the ghost entropy for the finite temperature
    density matrix (\ref{GHOSTDENSITY}), we find
    \[
        S_g = - \log (1+q) - {q\over 1+q}\,\log q = -S_s.
    \]
    We therefore see that the ghost carries negative entropy that exactly
    compensates that of the original spinor.  The ghost entropy is always
    between zero and $-1$, attaining the former in the pure state at zero
    temperature and the latter in the maximally mixed state at
    infinite temperature.

    We can also express the entropy in terms of the partition function by
    noting that for a finite temperature
    density matrix
    the definition (\ref{ENTROPYGHOST}) gives
    \begin{equation}
        S(\beta) = \left(1 - \beta\,{\partial\over\partial\beta}\right)\,\log
        Z.
        \label{ENTROPYZ}
    \end{equation}
    Cancellation of entropy between the spinor
    and ghost then follows trivially from
    (\ref{ENTROPYZ}) combined with the relation $1 = Z_s\,Z_g$.

    \section{$\sutwo$ and negative spin}

    In this section we will construct an irreducible representation of the
    $\sutwo$ Lie algebra \cite{ELLIOTT}
    on the ghost state space.  This representation
    has spin equal to $-\half$, and is pseudo-hermitian with respect to
    the indefinite inner product in the ghost state space.

    For later reference, let us write down the spin-$\half$
    representation
    \begin{eqnarray}
        J_+ &= &\left(
        \begin{array}{cc}
                        0 & 0 \\ 1 & 0
        \end{array}
        \right), \nonumber \\
        J_- &= &\left(
        \begin{array}{cc}
                        0 & 1 \\ 0 & 0
        \end{array}
        \right), \nonumber \\
        J_z &= &\left(
        \begin{array}{rc}
                        {\scriptstyle -\half} &  \\  &
                        {\scriptstyle\half}
        \end{array}
        \right). \nonumber
    \end{eqnarray}
    On the ghost state space,
    a set of generators hermitian with respect to the indefinite inner product
    (\ref{INNERPROD}) is given by
    \begin{eqnarray}
        J_x &\equiv &\phantom{-}{\textstyle {i \over 2}} \left( \sqrt{N+1}\, a -
        a^\dagger
            \sqrt{N+1}\right),
        \nonumber\\
        J_y &\equiv &- {\hhalf} \left( \sqrt{N+1}\, a + a^\dagger
            \sqrt{N+1}\right),
         \nonumber \\
        J_z &\equiv & N + {\hhalf}. \label{LIE}
    \end{eqnarray}
    It is easy to check that these generators satisfy the $\sutwo$ algebra
    \[
        [J_x, J_y] = iJ_z,\quad [J_y, J_z] = iJ_x,\quad [J_z, J_x] =
        iJ_y.
    \]
    We can define raising and lowering operators
    \begin{eqnarray}
        J_+ &\equiv &J_x + iJ_y = -i\,a^\dagger \, \sqrt {N+1}, \nonumber \\
        J_- &\equiv &J_x - iJ_y = \phantom{-} i\,\sqrt {N+1}\,a, \nonumber
    \end{eqnarray}
    satisfying $J_\pm^\dagger = J_\mp$ and
    \[
        [J_+, J_-] = 2 J_z.
    \]
    It is straightforward to calculate
    \[
        \mathbf{J}^2 = {\hhalf} \left( J_+ J_- + J_- J_+ \right) +
        {J_z}^2 = - {\textstyle {1 \over 4}}.
    \]
    Since $\mathbf{J}^2 = j(j+1)$, where $j$ denotes the spin, we
    see that
    \[
        j = -{\hhalf}.
    \]
    In other words, the representation that we have constructed
    has negative spin.

    Note that, since the entire state space is generated by
    applying the raising operator $J_+$ to the vacuum state, the
    representation is irreducible.  Also, in contrast to the positive
    spin irreducible representations, it is infinite-dimensional.

    However, a more useful definition of the dimension of a
    representation is given by the character of the identity
    $\chi(\mathbf{1})$.  In a later section, we shall see that for
    the ghost representation this is equal to $\half$.
    We will therefore argue that the dimension is effectively finite.

    The lowest weight state is the ghost ground state $\ket{\half}_g$, which has
    $J_z$ eigenvalue $m = 1/2$.
    As usual for a lowest weight state, it
    satisfies the relation $m=-j$.

    A peculiar feature of the representation is its asymmetry with
    respect to interchange of $J_-$ and $J_+$.  In particular,
    there is no highest weight state $\ket{-\hhalf}_g$ satisfying
    $J_+ \ket{-\hhalf}_g = 0$.

    It is straightforward to check that, in terms of the basis
    vectors (\ref{BASIS})
    we have
    \begin{eqnarray}
        J_+ \ket{m} &= &i\, \left(m+\hhalf\right)
                       \,\ket{m+\hhalf}, \qquad m = \hhalf, \textstyle{3\over
                       2}, \ldots
                         \nonumber \\
        J_- \ket{m+1} &= &i\, \left(m+{\hhalf}\right)
                       \,\ket{m}.
                \label{JPLUSACTION}
    \end{eqnarray}
    The factor $m + \half$ is the continuation to negative spin $j = -\half$ of the standard
    expression $\left[ (j + m + 1)(j -m)\right]^{\half}$
    \cite{ELLIOTT} from the representation theory
    of $\sutwo$.

    \section{Finite $\SUTWO$ transformations}

    In this section we will discuss the representation of finite
    $\SUTWO$ transformations on the ghost state space.
    Rather than exponentiating the $\sutwo$ generators by brute force,
    we will present a simple guess for the finite
    form of the transformations, prove that these indeed represent
    $\SUTWO$ transformations, and then relate them to the Lie algebra
    generators.

    Our first observation is that states in the combined spinor-ghost
    space of the special form
    \begin{equation}
        \left(
        \begin{array}{c}
                        1 \\ z
        \end{array}
        \right)
        \otimes
        \left(
        \begin{array}{l}
                        1 \\ z \\ z^2 \\ \vdots
        \end{array}
        \right)  \label{NORMALIZED}
    \end{equation}
    have unit normalization with respect to the indefinite inner product;
    in particular $(1 + zz^*) (1 - zz^* + (zz^*)^2 - \dots) = 1$.

    Now consider the effect of the $\SUTWO$ transformation
    \cite{ELLIOTT}
    \[
        \left(
        \begin{array}{cc}
                        a & b \\ c & d
        \end{array}
        \right) =
        \left(
        \begin{array}{rr}
                        \alpha & \beta \\ -\bar \beta & \bar
                        \alpha
        \end{array}
        \right).
    \]
    The spinor state becomes
    \begin{equation}
        (a + bz)\left(
        \begin{array}{c}
                        1 \\ z'
        \end{array}
        \right), \label{PARTICLESIDE}
    \end{equation}
    where $z'$ is given by the $CP^1$ conformal transformation
    \begin{equation}
        z' = {c + dz\over a + bz}. \label {CONFORMAL}
    \end{equation}
    Since the representation has to preserve the inner product, our ansatz
    is that the state after the rotation will be of
    the normalized form
    \begin{equation}
        \left(
        \begin{array}{c}
                        1 \\ z'
        \end{array}
        \right)
        \otimes
        \left(
        \begin{array}{l}
                        1 \\ z' \\ {z'}^2 \\ \vdots
        \end{array}
        \right).  \label{TRANSFORMED}
    \end{equation}
    If true, we see from
    (\ref{PARTICLESIDE}) that the transformed ghost state has to
    be
    \begin{equation}
        {1\over a+bz}
        \left(
        \begin{array}{l}
                        1 \\ z' \\ {z'}^2 \\ \vdots
        \end{array}
        \right)
        =
        {1\over a+bz}
        \left(
        \begin{array}{l}
                        1 \\ \left({c + dz\over a + bz}\right) \\ \left({c + dz\over a + bz}\right)^2 \\ \vdots
        \end{array}
        \right). \label{GHOSTSIDE}
    \end{equation}
    This will give us the information we need to determine the ghost
    $\SUTWO$ transformation matrices in terms of the parameters $a$, $b$,
    $c$ and $d$.  Indeed, we
    will first show that this transformation can be written in terms of
    a linear operator on the ghost state space.
    We will then verify that these linear operators represent
    $\SUTWO$ under
    composition.

    To exhibit the linear operator realizing
    the transformation (\ref{GHOSTSIDE}),
    we observe that, as long as $a\ne 0$ and
    $|z| < |a/b|$,
    we can expand the expression for the transformed ghost state in terms
    of positive powers of $z$.
    Doing this explicitly, we see that the expression (\ref{GHOSTSIDE}) is
    equivalent to the linear transformation
    \begin{equation}
        {1\over a}\cdot\left(
        \begin{array}{cccc}
            1 & -{b\over
            a} & \left({b\over
            a}\right)^2 & \dots \\
            {c\over a} &
            \left({d\over a}-{2bc\over
            a^2}\right) & \dots & \\
            \left({c\over a}\right)^2 & \dots & & \\
            \vdots & & &
        \end{array}
        \right)
        \left(
        \begin{array}{l}
                        1 \\ z \\ {z}^2 \\ \vdots
        \end{array}
        \right).  \label{EXPANSION}
    \end{equation}
    The expression for the general element
    of this candidate $\SUTWO$ matrix is easily calculated to be
    \begin{eqnarray}
        U_{ij} &=&
           \sum_{k=0}^{\min \left(i,j\right)} (-)^{j-k}\,{i + j - k \choose i}{i \choose k}
              \,a^{-i-j+k-1}\,b^{\,j-k}\,c^{\,i-k}\,d^{\,k},
              \label {GHOSTUNITARY}
    \end{eqnarray}
    for $i, j = 0, 1, 2, \ldots$

    To show that these linear operators $U$ represent
    $\SUTWO$ transformations under composition, it is enough to show that this is
    true on states of the special form
    \[
    \left(
        \begin{array}{l}
                        1 \\ z \\ z^2 \\ \vdots
        \end{array}
        \right),
    \]
    because these states linearly span the whole ghost state space.
    For these states, the effect of $U$ is by construction given by
    the expression (\ref{GHOSTSIDE}).
    It is readily verified that the conformal transformations
    $z \to \left({c + dz\over a + bz}\right)$ in (\ref{GHOSTSIDE})
    form a realization of
    $\SUTWO$ under composition.  For the prefactor in
    (\ref{GHOSTSIDE}), it is easy to check that, under
    composition of transformations parametrized by
    ${\scriptstyle{
        \left(
        \begin{array}{cc}
                        a & b \\ c & d
        \end{array}
        \right)}}
    $
    and
    ${\scriptstyle{
        \left(
        \begin{array}{cc}
                        a' & b' \\ c' & d'
        \end{array}
        \right)}}
    $,
    the prefactor becomes by linearity
    \[
        {1\over\left(a + bz\right)\left(a' + b' \left({c + dz\over a + bz}\right)\right)} =
        {1\over\left(a'a + b'c\right) + \left(a'b + b'd\right) z},
    \]
    as required.  This confirms that matrices of the form
    (\ref{GHOSTUNITARY}) represent $\SUTWO$ transformations.

    So far we have not been very careful with respect to the
    domains of the transformations $U_{ij}$.
    In fact,
    the domain of the linear operator defined by (\ref{GHOSTUNITARY})
    is not the entire state space.  This can be seen from the
    growth condition $|z| < |a/b|$ required for equivalence of
    (\ref{GHOSTSIDE}) and (\ref{GHOSTUNITARY}).
    Because the composition of two transformations of the form
    (\ref{GHOSTUNITARY})
    only makes sense if the range of the first overlaps
    with the domain of the second, matrices of the form (\ref{GHOSTUNITARY})
    cannot be arbitrarily composed.  In other words, the formal
    sums appearing in the matrix multiplication may not converge.
    Although one may be able to make sense of these formal sums by
    analytic continuation, doing this will be outside the scope of the
    present article.

    Also note that the representation
    matrix becomes singular when
    $a \to 0$, which happens for rotations by $\pi$ that
    take the unit vector $\mathbf{z}$ to $-\mathbf{z}$.
    This is a consequence of the fact, noted in the previous
    section, that the
    representation has no highest weight state, since these
    rotations would normally exchange lowest and highest weight states.

    The conclusion is that the Lie algebra representation of the
    previous section
    does not exponentiate to a give a representation of the full group.

    We can prove that the representation matrix $U$ is pseudo-unitary
    on its domain by showing that it conserves the norm of all states of the form
    \[
        \left(
        \begin{array}{l}
                        1 \\ z \\ {z}^2 \\ \vdots
        \end{array}
        \right),
    \]
    whose linear span is the entire state space.
    In particular, applying (\ref{GHOSTSIDE}), the norm of the
    transformed state becomes
    \begin{eqnarray}
      \lefteqn{
        \left|
        {1\over a + bz}
        \right|^2
        \left(
            1 -
            \left|
            {c + dz\over a + bz}
            \right|^2 +
            \left|
            {c + dz\over a + bz}
            \right|^4 + \dots
        \right)
      } \nonumber\\
        &= &{1\over |a + bz|^2 + |c + dz|^2} \nonumber \\
        &= &{1 \over 1 + z\bar z}, \nonumber
    \end{eqnarray}
    where in the last line we have used the fact that the $\SUTWO$
    matrix
    $
        \left(
        \begin{array}{cc}
                        a & b \\ c & d
        \end{array}
        \right)
    $
    is unitary, so that its columns are orthonormal.
    This is just the norm of the original state, which completes
    the proof.

    Note that pseudo-unitarity with respect to the indefinite inner product
    \[
        U^{-1} = \eta\, U^\dagger\,\eta.
    \]
    implies that the rows or columns of the matrix $U$ are
    orthonormal with respect to the inner product $\eta$.

    Finally, we can explicitly relate these finite transformations to the
    ghost Lie algebra generators we wrote down before.  Taking the
    one-parameter subgroup
    \[
        \left(
        \begin{array}{cc}
                        a & b \\ c & d
        \end{array}
        \right) =
        \left(
        \begin{array}{rr}
                        \cos {\theta\over 2} & \sin {\theta\over 2} \\
                        -\sin {\theta\over 2} & \cos {\theta\over 2}
        \end{array}
        \right),
    \]
    we can differentiate the corresponding ghost representation
    matrix in (\ref{GHOSTUNITARY}) to get the Lie algebra
    element
    \[
        {d\over d\theta}_{\theta = 0} U(\theta) = \hhalf
        \left(
        \begin{array}{ccccc}
                           & -1 &     &    & \\
                        -1 &    & -2  &    & \\
                           & -2 &     & -3 & \\
                           &    &  -3 &    & \ddots \\
                           &    &     & \ddots  &
        \end{array}
        \right)
    \]
    This is just our ghost $\sutwo$ generator
    $iJ_x = -{\textstyle {1 \over 2}} \left( \sqrt{N+1}\, a + a^\dagger
    \sqrt{N+1}\right)$.
    Similarly, the generators $iJ_y$ and $iJ_z$ are the Lie algebra generators
    corresponding respectively to
    \[
        \left(
        \begin{array}{rr}
                        \cos {\theta\over 2} & i\sin {\theta\over 2} \\
                        i\sin {\theta\over 2} & \cos {\theta\over 2}
        \end{array}
        \right)
    \]
    and
    \[
        \left(
        \begin{array}{rr}
                        e^{-i\,{\theta\over 2}} & 0 \\
                        0 & e^{i\,{\theta\over 2}}
        \end{array}
        \right).
    \]

    \section{Combining representations}

    In this section we study the product
    representation in the spinor-ghost system.
    We will discuss the
    sense in which the positive and negative spin representations cancel
    to give an effectively trivial representation.

    First, notice that there is a lowest weight state in the
    spinor-ghost product state space given by
    \[
        \ket{0}\equiv\ket{\textstyle
        -\half}_{s}\otimes\ket{\hhalf}_{g}.
    \]
    By definition, it is annihilated by
    $J_- \equiv J_-^s + J_-^g$.
    Using the relation
    \begin{eqnarray}
        \mathbf{J}^2 &=& J_+ J_- + J_z^2 - J_z\nonumber
    \end{eqnarray}
    it trivially follows that
    \[
        \mathbf{J}^2 \, \ket{0} = 0.
    \]
    We are therefore dealing with an $\sutwo$ representation of
    spin $0$ in the product space.  What is unusual about this
    representation, however, is that $J_+ \ket{0} \ne
    0$.  However, this state has zero norm.
    In fact, this is true for all higher states in the ladder since
    \[
        J_+^{n} \ket{0} = i^n \,n!\sqrt 2 \, \ket{n},
    \]
    where $\ket{n}$ is defined in (\ref{NTILDEN}) and is null.
    In general, we would like to truncate the ladder of states
    generated by $J_+$ as soon as we reach a null state.

    In section (3), we chose the BRST operator $Q$ equal to $\mathbf{J}^2$.
    In terms of the basis (\ref{NTILDEN}), it is not hard to
    calculate
    \[
        \mathbf{J}^2 \ket{\tilde n} = 2n\ket{n}, \qquad \mathbf{J}^2 \ket{n} = 0,
    \]
    leading to the matrix representations (\ref{Q1}) and (\ref{Q2}).
    Since $\mathbf{J}^2$ is the Casimir operator, the generators
    $J_i$ all commute with $Q$.  In other words, they are physical
    operators in the sense of section 3.

    We can therefore consistently reduce these operators to the
    cohomology of $Q$,
    and use (\ref{AREDUCED})
    to define the induced
    operators  $[J_+]$, $[J_-]$ and $[J_z]$ on the quotient space
    by
    \[
      [J_i] \,(\ket{\phi} + \im Q) = J_i \ket{\phi} + \im Q.
    \]
    The cohomology consists of the single class $\ket{0} + \im Q$,
    and is a one-dimensional, positive definite Hilbert space.
    The induced generators $\{[J_+], [J_-], [J_z]\}$
    are zero, corresponding to the trivial
    representation of $\sutwo$.

    \section{Character and dimension formulae}

    The characters of the positive spin representations of $\SUTWO$
    are given by \cite{ELLIOTT}
    \[
        \chi^{(j)}(\theta) = {\sin \left(j + \half\right)\,\theta \over \sin
        \half\theta},
    \]
    and satisfy the orthogonality relation
    \begin{equation}
        {1\over 2\pi} \,\int_0^{2\pi}
        d\theta\,\chi^{j_1}(\theta)\,
        \chi^{j_2}(\theta) \, (1 - \cos \theta) = \delta_{j_1 j_2}.
        \label{ORTHOGONALITY}
    \end{equation}
    We remind the reader that a character is a function on the conjugacy
    classes of a group.  For $\SUTWO$,
    all rotations through the same angle $\theta$ are in the same class,
    irrespective of the direction of their axes.

    We will now attempt to define a meaningful character for the spin $(-\half)$
    representation.

    In accordance with the discussion of the previous section, we
    would like the character of the product representation to
    be equal to that of the trivial representation.
    The expression for the character should therefore
    be invariant under the BRST reduction used to eliminate
    the null states in the product representation.
    An expression that ignores the discarded states is given by
    \[
        \chi^{(-\half)\otimes(\half)}(\theta) = \Tr (\mathbf{1}_s \otimes\eta) \,e^{i\theta
        J_z},
    \]
    where $\eta = (-)^{N_g}$ and $e^{i\theta J_z}$
    is the representative of the conjugacy class corresponding to angle
    $\theta$.  Here
    $J_z = J_z^s\otimes \mathbf{1}_g + \mathbf{1}_s\otimes J_z^g$.

    From the properties of the trace, it immediately follows that
    \[
        \chi^{(-\half)\otimes(\half)}(\theta) =
        \chi^{(-\half)}(\theta) \, \chi^{(\half)}(\theta),
    \]
    where
    \begin{eqnarray}
       \chi^{(-\half)}(\theta) &= &\Tr \eta \,e^{i\theta
          J_z^g},  \nonumber \\
       \chi^{(\half)}(\theta) &= &\Tr e^{i\theta
          J_z^s}.  \nonumber
    \end{eqnarray}
    Furthermore, we have
    \begin{eqnarray}
        \chi^{(-\half)}(\theta) &= &\Tr \eta \,e^{i\theta J_z} \nonumber\\
           &= &e^{i\theta/2} \left(1 - e^{i\theta} + e^{2i\theta} - \dots\right)
           \nonumber\\
           &= &{1\over 2 \cos {\half \,\theta}}, \label{CHARACTER}
    \end{eqnarray}
    which is just the inverse of the
    character
    \[
        \chi^{(\half)}(\theta) = {\sin \theta \over \sin
        \half\theta} = 2 \cos {\hhalf \,\theta}
    \]
    of the spin $\half$ representation.  This means that
    \[
        \chi^{(-\half)}\chi^{(\half)} = 1 = \chi^{(0)}.
    \]
    This is consistent with the fact that the product of the
    spinor
    and the ghost gives an effectively trivial representation.

    The character $\chi^{(-\half)}$ is not
    normalizable with respect to the measure in (\ref{ORTHOGONALITY}).
    Indeed, because of the singularity at $\theta = \pi$, it does not live
    in the same function space.  However,
    we can take its inner product with positive spin characters,
    which means that we may regard it as a distribution.

    Note that the singularity at $\theta = \pi$ is to be expected.  Indeed,
    we have seen already that certain elements in the conjugacy
    class of rotations by $\pi$ are ill-defined (notably the finite
    rotations in section 8 where $a\to 0$).  The singularity in
    the character reflects this fact.

    The dimension of a representation is
    conventionally related to the character as $d = \chi({\mathbf{1}})$.  For the
    ghost, this gives a fractional dimension
    \[
      d = \half.
    \]
    This is consistent with the
    argument that the combined spinor-ghost system effectively
    transforms in the trivial representation of $\SUTWO$,
    which has dimension $1$.  Indeed, the dimension of a product
    representation is the product of the dimensions, so that the
    combined system will have effective dimension $\half\cdot 2 =
    1$, which is indeed correct.

    The dimension can also be related to the entropy as
    follows.  Conventionally, the entropy of a maximally mixed
    state is related to the dimension of the state space $\mathcal{H}$ by
    \[
        S_{\mathrm{extremal}} = \log_2 \dim \left(\mathcal{H}\right).
    \]
    For the ghost, we have seen that $S_{\mathrm{extremal}} = -1$,
    consistent with a dimension $\dim \left(\mathcal{H}\right) = \half$.

    \section{Conclusion}

    In this article, we studied the properties of a BRST ghost
    degree of freedom complementary to a two-state spinor.
    We showed that the ghost entropy
    ranges between $0$ and $-1$ and cancels the entropy of the
    spinor.

    We then constructed an irreducible
    representation of the $\sutwo$ Lie algebra
    of negative spin, equal to $-\half$, on the ghost state
    space.  We were also able to exhibit representation matrices
    for finite $\SUTWO$ transformations, although we found that
    these become singular for certain $\SUTWO$ elements, and cannot be
    arbitrarily composed because of domain issues.
    In other words, the Lie algebra representation does not
    exponentiate to a give a representation of the whole group.

    Since we chose the BRST operator $Q$ to be equal to the
    Casimir operator $\mathbf{J}^2$ of the product representation, the
    generators of the product representation all commute with $Q$.
    Using this, we discussed the sense in which the positive and negative spin
    representations combine  to
    give an effectively trivial representation in the product space
     and showed that a
    character can be defined for the negative representation in a
    way that is consistent with this cancellation.
    We argued that a fractional effective dimension of $\half$ can be
    assigned to the ghost representation.

    The methods of this article can be expanded to the study of arbitrary
    negative spin representations of $\sutwo$.
    Work in this direction is in progress \cite{MYSELF}.

\end{document}